# Coherent Optical Clock Down-Conversion for Microwave Frequencies with 10⁻¹⁸ Instability

Takuma Nakamura[1]*, Josue Davila-Rodriguez[1], Holly Leopardi[1,2], Jeff A. Sherman[1], Tara M. Fortier[1,2], Xiaojun Xie[3], Joe C. Campbell[3], William F. McGrew[1,2], Xiaogang Zhang[1], Youssef S. Hassan[1,2], Daniele Nicolodi[1,2], Kyle Beloy[1], Andrew D. Ludlow[1,2], Scott A. Diddams[1,2] and Franklyn Quinlan[1,2]†.

[1]Time and Frequency Division, National Institute of Standards and Technology, 325 Broadway, Boulder, CO 80305, USA
[2]Department of Physics, University of Colorado Boulder, 440 UCB Boulder, CO, 80309, USA
[3]Department of Electrical and Computer Engineering, University of Virginia, Charlottesville, VA, 22904, USA
*takuma.nakamura@nist.gov; †fquinlan@nist.gov

**Abstract:** Optical atomic clocks are poised to redefine the SI second, thanks to stability and accuracy more than one hundred times better than the current microwave atomic clock standard. However, the best optical clocks have not seen their performance transferred to the electronic domain, where radar, navigation, communications, and fundamental research rely on less stable microwave sources. By comparing two independent optical-to-electronic signal generators, we demonstrate a 10 GHz microwave signal with phase that exactly tracks that of the optical clock phase from which it is derived, yielding an absolute fractional frequency instability of $1\times10^{-18}$ in the electronic domain. Such faithful reproduction of the optical clock phase expands the opportunities for optical clocks both technologically and scientifically for time dissemination, navigation, and long-baseline interferometric imaging.

Motivated by placing tighter constraints on physical constants and their possible variations (*1, 2*), precise measurements of gravitational potential (*3*), and gravitational wave detection (*4*), the accuracy and stability of optical clocks have continued to improve, such that they now outperform their microwave counterparts by orders of magnitude. These clocks – based on optical transitions in atoms and ions such as ytterbium, strontium, and aluminum (*5, 6*) – take advantage of an operating frequency that can exceed 1000 THz. This corresponds to sub-dividing a second into mere femtoseconds, allowing for an extremely precise measurement of time. The frequency stability of an optical clock, best described as the fraction of the clock's frequency fluctuations relative to its nominal operating frequency, can now reach below $10^{-18}$ (*3*). Perhaps more important, optical clocks can reach $10^{-16}$ performance in a matter of seconds, rather than the month-long averaging required of a microwave cesium fountain clock, which currently defines the SI second, to reach this level (*7*). With such extraordinary performance, in conjunction with optical clocks meeting other benchmarks laid out by the International Committee for Weights and Measures (CIPM) (*8*), a redefinition of the SI second appears inevitable.

The range of applications enjoyed by optical clocks can be further extended by transferring their stability to the electronic domain. Doppler radar sensitivity, particularly for slow moving objects, is strongly determined by the frequency noise of the transmitted microwaves and could see a large sensitivity enhancement by using optically derived electrical signals (*9-11*). Astronomical imaging and precise geodesy with very long baseline interferometry (VLBI) also rely on highly frequency-stable electronic sources (*12*). In ground-based VLBI, microwave and mm-wave signals are detected at receivers spread across the globe and are coherently combined to form exceptionally high-resolution images of cosmic objects. Moving to a space-based VLBI network greatly increases the resolution (*13*) and avoids atmospheric distortions that limit the observation time. In spaced-based VLBI, maintaining phase coherence with electronic local oscillators that have optical clock level stability could increase observation time from seconds to hours with a commensurate increase in the number of objects that can be imaged with high fidelity.

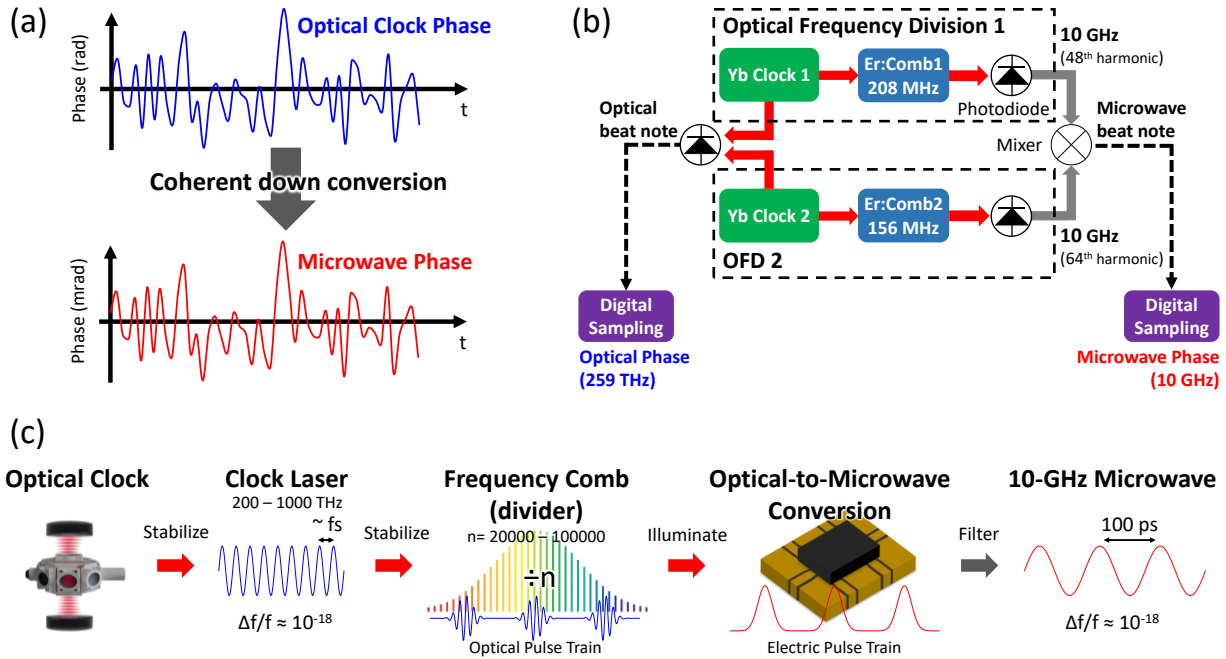

**Fig.1.** Coherent optical clock down conversion (a) The optical clock phase is transferred to the microwave domain with fluctuations scaled by the optical-to-microwave frequency ratio. (b) Simplified setup of phase and frequency stability measurements. The outputs of two independent Yb optical atomic clocks generate microwave signals at 10 GHz. By frequency-mixing the 10 GHz outputs, the relative phase fluctuations are recorded. A direct optical beat note reporting the relative optical phase of the Yb clocks is also recorded. (c) Schematic of microwave generation from an optical atomic clock. An optical frequency comb is stabilized to the optical clock laser. Optical-to-microwave conversion via high-speed photodetection generates a train of electrical pulses. Selectively filtering the electrical signal results in a microwave tone phase-coherent with the optical clock.

With the use of fiber-based optical frequency combs and state-of-the-art photodetectors, we have generated and evaluated microwave signals that preserve the phase of the optical clocks from which they are derived with sub-femtosecond precision. The resulting frequency stability on a 10 GHz carrier is better than any other microwave source, and represents a 100-fold stability improvement over the best Cs fountain clocks. Moreover, the inaccuracy of the optical-to-microwave frequency conversion was measured to be less than $1\times10^{-19}$. Preservation of the optical clock phase opens up the possibility of distant optical clock synchronization with microwave carriers for applications in navigation and fundamental physics. Lastly, coherently linking an optical atomic frequency standard to the electronic domain allows for future calibration of electronic clocks – an important consideration for the redefinition of the SI second based on an optical atomic transition.

Generating an electronic signal linked to an optical clock is the physical implementation of dividing the optical clock frequency by a large integer (*10*). The concept is shown in Fig. 1. The first element in this division process is the optical frequency comb (OFC) – a laser source consisting of an array of discrete, evenly spaced frequency tones that span 100s of THz (*14, 15*). When an OFC is locked to a clock, each individual tone of the comb carries the same frequency stability as that of the master clock. (Transferring the clock stability to each line is nearly perfect – added instabilities are only at the $10^{-20}$ level or below (*16-18*).) The broad spectrum of the comb gives rise to a train of optical pulses with sub-picosecond duration. The repetition rate of these pulses, typically in the range of 10s of MHz to a few GHz, is coherently linked to the optical clock frequency, but is divided down to a much lower microwave frequency. Importantly, the clock frequency fluctuations also divide, such that the fractional frequency stability is maintained. Thus locking an OFC to an optical clock operating at 259 THz and fractional frequency instability of $10^{-16}$ can produce a 100 MHz pulse train whose repetition fractional frequency instability is also $10^{-16}$.

The repetition rate of an OFC is accessible with electronics, such that by illuminating a high-speed photodiode can, in principle, create a train of electrical pulses with optical clock stability. Optical-to-electrical conversion that preserves optical clock-level stability is not straightforward, however, for this process must contend with the photodiode's nonlinear response engendered by the high peak intensities of ultrashort pulses, the quantum limits of light detection, and the vagaries of electron transport dynamics (*19-21*). Considerable effort has been devoted to understand and overcome the limitations of photodiodes for optical-to-electrical conversion of ultrastable optical pulse trains, leading to new detector designs (*22*), and techniques to lower the impact of quantum noise in the phase stability of the optically derived electronic signal (*23*). This progress has set the stage for the demonstration of electrical signal generation that faithfully reproduces the frequency and phase of a state-of-the-art optical clock.

With frequency stability better than any other microwave source, measurements required constructing two systems and comparing them against one another. A simplified schematic diagram of the microwave generation and measurement is shown in Fig. 1(b). Ten GHz microwaves were derived from two independent ytterbium optical lattice clocks, each of which demonstrate state-of-the-art stability (*3*), and absolute frequency verified against the SI second (*24*). The OFCs were based on two home-built erbium fiber mode-locked lasers with respective repetition rates of 208 MHz and 156 MHz. These OFCs were engineered for long-term, phase-slip-free operation, and contribute negligible excess noise. The optical pulse trains from the OFCs were detected with photodiodes designed for high speed and high linearity (*22*), from which electrical pulse trains were generated. The frequency spectrum of these electrical pulses is an array of tones at the harmonics of the pulse repetition rate. Narrow electrical bandpass filters selected a single frequency near 10 GHz from each system for evaluation. As the repetition rates of the two lasers are not the same, the nominal 10 GHz outputs represented the 48th harmonic and 64th harmonic of the respective systems. These 10 GHz outputs were combined in a microwave frequency mixer, producing a difference frequency near 1.5 MHz that was digitally sampled (*25*), from which the microwave phase was extracted. From the phase, frequency stability and accuracy were determined. In addition to the comparisons of the microwave outputs, a direct optical comparison of the clocks was made. This was performed by combining the optical clock signals onto a single photodetector, directly generating an electrical signal at an intentionally offset beat frequency between the clocks. This beat frequency was also digitally sampled and recorded. Comparing the phase of the difference frequency of the microwave outputs to that of the optical beat frequency allowed for confirmation of high-fidelity phase and frequency transfer to the electronic domain. More details of the experimental setup may be found in the supplement.

It is interesting to note the level of resolution required to measure a fractional frequency instability of $10^{-16}$ at 1 second on a 10 GHz signal. This implies tracking phase changes corresponding to only one millionth of a cycle. (For comparison, measuring $10^{-16}$ on an optical carrier requires splitting the cycle by a factor of about 40.) Achieving this level of phase resolution, and maintaining it over several hours, was accomplished in the following ways. First, we utilized microwave amplifiers with low flicker noise, and we routed signals with temperature-insensitive cabling. This gave us an output with significant power (~10 mW) without sacrificing stability. Second, by frequency-mixing the 10 GHz outputs we shift the measurement to a 1.5 MHz carrier. This reduces the requirements on the fractional stability we must measure by ~7000 (10 GHz/1.5 MHz). Another advantage of this measurement scheme is the high phase resolution is achievable without requiring the two sources to oscillate at exactly the same frequency. This gives our measurement system some dexterity in comparing high stability signals from independent sources.

Optical and microwave phase fluctuations, continuously recorded over 44,000 seconds, are shown in Fig. 2. In the Fig. 2(a), the phase fluctuations of the optical clock have been scaled by a factor equal to the optical-to-microwave frequency ratio (nearly 26,000) to illustrate the extremely high fidelity in the optical-to-microwave transfer. The relative phase can also be expressed as a timing fluctuation, and is bounded by ± 30 femtoseconds for both optical and microwave signals. Also shown in Fig. 2(a) is the point-by-point difference between optical and microwave measurements, limited to root mean square (rms) fluctuations of 60 microradians, corresponding to a rms relative timing fluctuation of only 900 attoseconds. This implies that optical clocks with even higher stability can be converted to microwave signals without loss of fidelity. The strong correlation between the optical and microwave phases is shown in Fig. 2(b). The degree of correlation is quantified by the correlation coefficient, ranging from 0 for completely uncorrelated phases to a maximum value of 1 for complete linear correlation between optical and microwave phase (*26*). The calculated correlation coefficient for the data in Fig. 2 is 0.998. Such femtosecond-level, high coherence optical-to-microwave conversion opens up the possibility of connecting distant optical clocks with a microwave link. Currently, these clocks can be linked optically through fiber or over free space (*27*), enabling state-of-the-art clock comparisons and synchronization. Free-space optical links are particularly useful for the many situations where a dedicated fiber link is not available, but can become ineffective due to poor weather or dusty conditions. The lower loss of microwave transmission could prove advantageous under such conditions by providing a link that would be impossible to maintain optically.

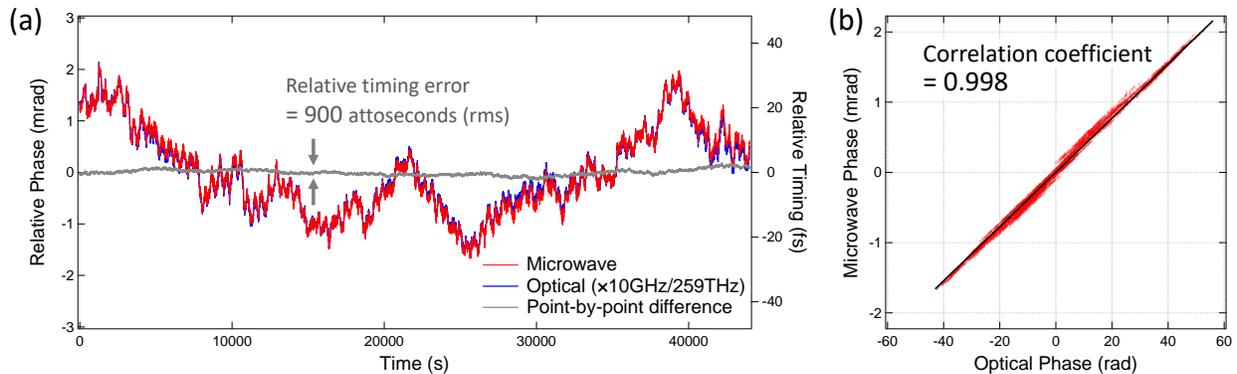

**Fig.2.** Optical and microwave phase, timing, and relative coherence. (a) Phase and corresponding timing fluctuations in optical and microwave domains. The point-by-point difference in the (scaled) optical and microwave phase records is shown in grey. (b) Phase correlation plot demonstrating a correlation coefficient of 0.998. The black line is the expected slope given by the optical-to-microwave frequency ratio.

Whereas fluctuations in the phase provide all the frequency and timing stability information of an oscillator, the fractional frequency instability is the more typical performance benchmark. Figure 3 displays the fractional frequency instabilities derived from the same 44,000s duration phase measurements shown in Fig. 2(a). The frequency stability of the derived microwaves followed that of the optical clocks precisely, ultimately yielding an absolute fractional frequency instability of $1\times10^{-18}$. This is 100 times more stable than the Cs fountain clocks that currently serve as the best realization of the SI second. The short-term stability also exceeds that of other microwave sources, the best of which are microwave oscillators based on whispering gallery mode resonances in cryogenically cooled sapphire (*28*). There are several known techniques for improving optical clock performance beyond that which is demonstrated here, such as real-time blackbody-shift corrections (*3*), zero dead-time operation (*29*), and high-performance laser local oscillators (*30*), that have led to lower instabilities as indicated in the purple line of Fig. 3. Separate measurements of the added instability due to noise in our optical-to-microwave transfer, shown in pink in Fig. 3, reach $5\times10^{-17}$ at 1 second and $1\times10^{-18}$ at 200s. This indicates our optical-to-microwave down-conversion can support the highest stability optical clocks yet demonstrated without degradation.

In addition to stability, we examined possible frequency offsets in the optical-to-electrical transfer that would degrade the accuracy of the resulting microwave signal. This is best analyzed by comparing the separation in the Yb clock frequencies as determined by the microwave measurement and as determined by the direct optical beat. Table 1 shows the results of our accuracy analysis, and includes directly measured frequency differences without accounting for known systematic shift mechanisms in the clock systems (such as blackbody radiation-induced shifts). Both optical and microwave measurements yielded a fractional frequency offset near $-5.9\times10^{-17}$, consistent with the known offset between the Yb clocks used in our experiments. More importantly, the difference in the offset from microwave and optical measurements, again represented as a fractional offset, was only $2.5\times10^{-20}$. This is smaller than the statistical uncertainty of $9.6\times10^{-20}$ of the point-by-point difference shown above in Fig. 2. Thus any unintentional offsets resulting from the frequency transfer from the optical to microwave domain are well below the $\sim10^{-18}$ accuracy level of a state-of-the-art optical clock (*3, 31*).

Transferring the phase, the frequency stability, and the accuracy of optical clocks to the electronic domain has resulted in 100-fold improvement over the best microwave sources. With microwave signals having optical clock stability, one can envision a robust, phase-coherent system of ultrastable electronic signals capable of supporting future radar, communications, navigation, and basic science. Moreover, with residual instability of the optical-to-microwave link below that of the best optical clock demonstrations to date, further improvements to the absolute stability of microwaves can be expected.

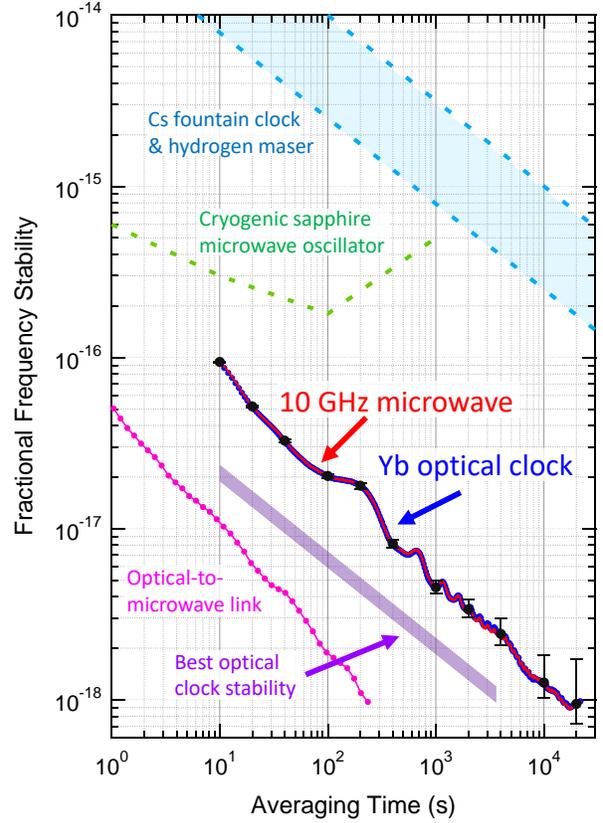

**Fig.3.** Fractional frequency stability comparison of state-of-the-art sources (*7, 28-30*). The microwave and optical signals locked to the Yb clocks are reported in terms of the Total Allan deviation, with error bars representing 1σ confidence intervals. (For all plots, the frequency stability is given by the Allan deviation.) Additionally, the residual instability of the optical-to-microwave link is well below state-of-the-art optical clocks.

|  | Frequency offset (Yb1 - Yb2) at 259 THz (1156 nm) | Fractional frequency offset |
|---|---|---|
| Optical measurement | -0.0152862 Hz | $(-5.8986\pm0.095)\times10^{-17}$ |
| Microwave measurement | -0.0152926 Hz | $(-5.9011\pm0.096)\times10^{-17}$ |
| difference | 0.0000064 Hz | $(2.5\pm9.6)\times10^{-20}$ |

**Table.1.** Frequency accuracy evaluation of the optical-to-microwave link. The optical clock frequency offsets were measured both optically and on the derived 10 GHz microwaves, then compared. The difference in the two measurements is consistent with zero at the $10^{-19}$ level.

**Acknowledgments:** We thank C. W. Oates for discussions about overall experiments and comments on the manuscript. We further thank J. C. Bergquist, D. Slichter, and L. C. Sinclair for their comments on this manuscript. **Funding:** This paper is supported by National Institute of Standards and Technology and DARPA. **Author contributions:** T.N., J.D., J.A.S., and F.Q. performed the microwave measurements. T.N., J.D., H.L., T.M.F., S.A.D., and F.Q. developed the Er:fiber frequency combs. X.X. and J.C.C. designed and fabricated the high-speed photodetectors. W.M.F., X.Z., Y.S.H., D.N., K.B., and A.D.L. designed, constructed, and operated the Yb optical clocks. F.Q. supervised the work. All authors contributed to the final manuscript. **Competing interests:** The authors declare no competing interests. **Data and materials availability:** All the data is available upon reasonable request from the corresponding authors.

# Supplementary Text:

Yb optical lattice clocks and fiber links

This section provides more details of the Yb optical lattice clocks and how two Yb clock frequencies were delivered to a separate lab which housed the optical frequency combs (OFCs). The experimental setup is shown in Fig. S1. The two labs are connected by two 80-m actively noise cancelled fiber links (*32, 33*). By sending a single Yb clock signal through both links simultaneously, the stability of the 80 m fiber links was evaluated at $3\times10^{-18}$ at 10 s, reducing to mid-$10^{-20}$ after a few hundred seconds averaging. Given the known Yb clock stability, the instability of the links was deemed not to impact the fidelity of the clock signal delivered to the OFC lab. The distributed clock light was a continuous-wave (CW) 1156 nm laser that was pre-stabilized to an ultra-low-expansion (ULE) cavity by the Pound–Drever–Hall technique (*34*). This single laser was used for both Yb systems, with clock frequency corrections implemented on separate light paths. To spectroscopically probe the Yb atoms of both systems, the CW laser was frequency doubled by a waveguided periodically poled lithium niobate (PPLN) crystal, and then delivered to the trapped Yb atoms. The frequency correction signal from one of the Yb clocks (Yb 1 in Fig. S1) was fed back to an acousto-optic modulator (AOM1 in Fig. S1) that was placed at the branch for the cavity. AOM1 was not used for stabilization to the cavity, rather it added a frequency offset which was modulated by the feedback signal from Yb1. In this way the output from the laser was stabilized to Yb1. Therefore, the fiber link from Yb1 to the first frequency comb (fiber link 1 to Comb 1) was also stabilized to Yb1. Additional acousto-optic modulators (AOM3 and AOM5) independently overwrote the feedback signal from Yb1 and tuned the output frequency to stabilize to the second Yb clock (Yb2). Note that feedback signals for AOM5 were halved compared with AOM3 because the optical frequency was not doubled on AOM5. Furthermore, AOM5 has two functions, fiber noise cancellation and stabilization to Yb2. Thus, the two Yb clocks were operated independently while served by a common CW laser. For short time scales, the reliance on a common optical cavity correlates the noise and stability of the two clocks. The feedback rate of the independent Yb clocks provide independent outputs for averaging times greater than a few seconds. To ensure we are comparing independent sources, the main text reports fractional frequency instability only for time scales longer than 10 s.

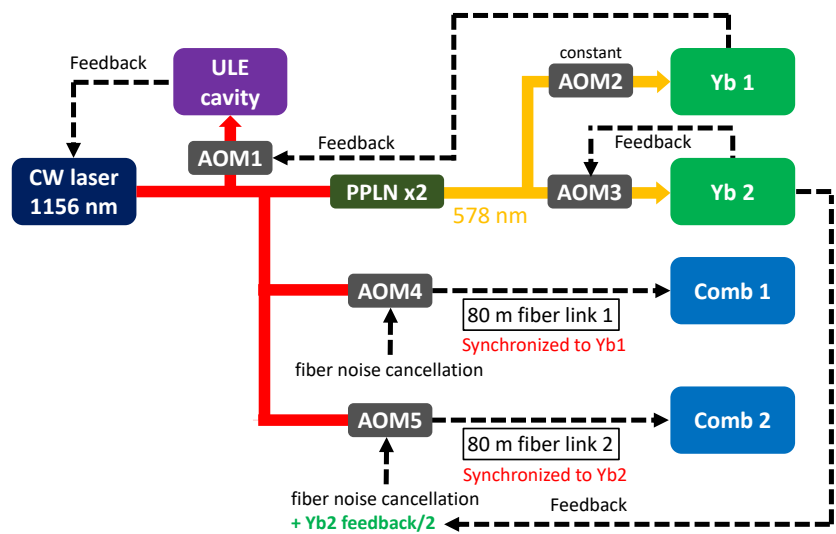

**Fig.S1.** Schematic diagram of Yb clocks and 80 m fiber links. A single CW laser operating at 1156 nm was pre-stabilized to a ULE cavity and used for probing two Yb systems. Feedback to AOM1 tuned the frequency of the CW laser to Yb1. AOM3 and AOM5 overwrote the feedback for Yb2. AOM4 and AOM5 are used for fiber noise cancellation over the 80 m link. AOM2 was only used to add a constant frequency offset. AOM, acousto-optic modulator; ULE, ultralow expansion; PPLN, periodically poled lithium niobate.

The two Yb clocks were operated by unsynchronized Ramsey spectroscopy with a free-evolution time of 400 ms. Cycle times of Yb 1 and Yb2 were 844 ms and 962 ms, respectively. Therefore, the interrogation timing of the two clocks continuously shifted. Small temperature fluctuations induced a varying blackbody radiation shift, worsening the long-term stability of each clock. Although this effect can be subtracted after processing (*3*), the correction was not applied here.

Er:fiber frequency combs

A schematic of the frequency combs used is shown in Fig. S2. Two home-made Er:fiber mode locked lasers were built for ultrastable microwave generation. Technical details of the two combs are almost identical with the exception of the repetition rate. The mode-locking scheme for these lasers was nonlinear polarization rotation, and an intracavity piezo electric transducer (PZT) and electro optic modulator (EOM) were implemented for repetition rate control (*35*). The output power from each mode-locked laser was approximately 50 mW with an optical spectral width of ~ 50 nm. The optical pulse train generated by each mode-locked laser was amplified by a polarization-maintaining (PM) Er-doped fiber amplifier (EDFA) and spectrally broadened by a PM highly nonlinear fiber (HNLF). Note that these combs have only one fiber-branch (so called "single-branch" configuration) for offset frequency, optical beat, and repetition rate detection. This method, as opposed to a multi-branch configuration (*36*), provided a simple way to achieve high stability, that otherwise requires more elaborate methods to reduce the differential noise for multiple fiber paths (*16-18, 37, 38*). Still, the optical path length of all branches consisting of free space optics after supercontinuum generation are set as short as possible and are enclosed to protect from air path fluctuations. The longer wavelength portion of the spectrum was frequency-doubled by PPLN for self-referenced offset frequency detection. Additionally, part of the supercontinuum centered at 1156 nm was extracted to obtain an optical beat note with one of the Yb clock frequency references. The signal-to-noise ratios of the offset frequency and optical beat detection were both approximately 40 dB in 100 kHz resolution bandwidth. Through a microwave frequency mixer, the optical beat note was mixed with the offset frequency, and phase locked to 126.5 MHz signal referenced to a hydrogen maser. This stabilized the repetition rate of the comb to the optical clock by feeding back to the EOM and the PZT for fast/small dynamic range and slow/large dynamic range control, respectively (*20, 39*). In practice, the offset frequency was loosely stabilized with a low gain and low bandwidth lock to keep frequency within the passband of a microwave bandpass filter. For microwave generation, pulses from the supercontinuum with wavelengths >1200 nm were directed to a high-speed photodetector. The photodetectors used were modified uni-traveling carrier (MUTC) photodiodes, specially designed for high power handling and high linearity (*22, 40*). In ultrastable microwave generation derived from optical pulses, the important nonlinearity is the conversion of power fluctuations on the optical pulse train to phase fluctuations of the derived microwave (*41, 42*), so-called AM-to-PM conversion. By operating at the appropriate "null" photocurrent, this nonlinearity in MUTC detectors was suppressed by more than 40 dB (*21*). The value of the null photocurrent depends on the applied bias voltage, pulse energy and detector linearity. For the results presented in the main text, the null point was around 0.6 mA, yielding a microwave power near -40 dBm. Additionally, an acousto-optic modulator was implemented for active intensity control to maintain null point operation for extended periods of time. This has the added benefit of reducing the intensity noise on the derived microwave signal. Thanks to the high linearity of the MUTC detector and intensity stabilization, the phase noise originating from optical power fluctuations was estimated -150 dBc/Hz at 1 s, corresponding to a fractional frequency instability of $1\times10^{-18}$ at 1 s.

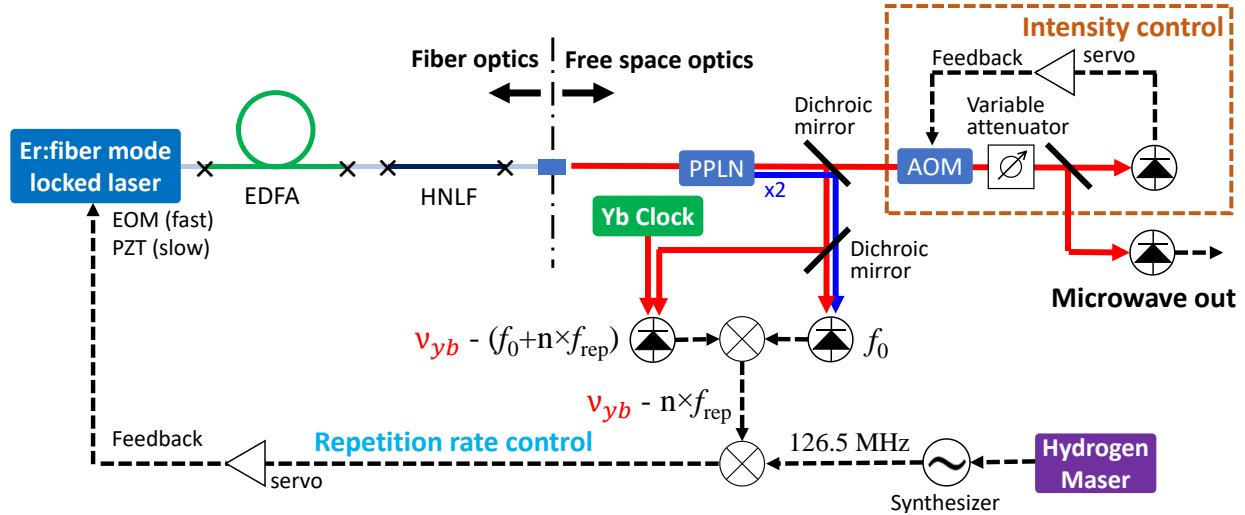

**Fig.S2.** Schematic diagram of one of the Er:fiber combs and microwave generation. The two combs are almost identical. Er:fiber lasers were mode-locked by nonlinear polarization rotation. An octave spanning spectrum was achieved by a polarization-maintaining Er-doped fiber amplifier (PM-EDFA) and a polarization-maintaining highly nonlinear fiber (PM-HNLF). The OFC repetition rates were stabilized to the Yb clocks. An intensity servo was implemented for minimization of the photodetector's amplitude-noise-to-phase-noise conversion. PPLN, periodically poled lithium niobate; EOM, electro-optic modulator; PZT, piezo-electric transducer.

Microwave measurement

Figure S3 shows the details of our microwave measurement setup. Evaluating the frequency stability and phase noise of our optically derived microwaves requires comparison against another source known to be of equal or better stability. This was achieved by comparing the output of the two independent 10 GHz signals in a frequency mixer. The resulting ~1.5 MHz signal was then evaluated for phase noise and frequency stability. Measurements of both the optical and microwave phase were performed via digital signal processing with software defined radio (SDR) (*25*). In SDR, input signals are first sampled by an analog-to-digital converter (ADC). They are then compared to a numerically controlled oscillator to extract phase and amplitude information of the input through IQ demodulation. Both the numerically controlled oscillator and the ADC are synchronized to a 10 MHz "master clock" reference signal. For stability measurements, it is critical that the reference signal stability be better than the signal under test. In anticipation of obtaining 10 GHz signals with relative instability below $1\times10^{-16}$ at 1 s, we opted in favor of an optically derived 10 MHz reference. This was based on the fact that, despite the leveraged gain from transferring the noise at 10 GHz to a 1.5 MHz carrier, the H-maser signal available to us would have skewed our residual noise and stability measurements. The optically derived 10 MHz was synthesized from the 2nd harmonic of comb1 at 416 MHz. This signal was used to clock a direct digital synthesizer (DDS), generating the required 10 MHz output. It is important to note that the 10 MHz generation was the result of a frequency division process, distinct from the frequency translation of the 10 GHz signals by the mixer. In frequency division, the noise of the signal is divided along with the carrier frequency. This resulted in negligible common-mode rejection of the 10 GHz noise, as described more fully in the following section.

A common CW 1070 nm laser was used to determine the stability of our optical-to-microwave conversion system. In this case, noise of the CW laser was cancelled out, leaving only the added noise in the down-conversion, electrical amplification, mixing, and noise measurement processes. To further isolate the noise added by the optical frequency comb and photodiode, the 10 GHz signals from each detector were separated into two identical measurement chains. In this case, the measurement technique of cross-correlation can be applied (*43*), which compares signals from the two chains and extracts correlated information between the two chains. Therefore, uncorrelated noise coming from the individual measurement chains, including amplifiers, mixers and SDRs, is eliminated by multiplying the Fourier transform of the temporal phase record of the two chains and averaging. Note that to obtain a high suppression of uncorrected phase noise, a large number of averages is necessary. As a consequence, the total measurement time is substantially increased. Single-chain and cross-correlation measurements of the phase noise power spectral density for the optical to microwave conversion are shown in Fig. S4(a). The bright tones at low frequency in the single-chain measurements are considered measurement artifacts, as they are known to originate in the SDR. The red trace represents the phase noise after cross-correlation. With 129 averages, approximately 5 to 10 dB suppression was obtained over the whole spectrum range, and the bright tones from the SDRs disappear. This result indicates the that the measurement system, including amplifiers and mixers, is the dominant noise source in the single chain measurements. To the best of our knowledge, the phase noise of -126 dBc/Hz at 1 s of the cross-correlation trace achieved here is the lowest level yet reported for a single system noise of an optical-to-microwave converter. In contrast, single chain measurement shows -120 dBc/Hz at 1 s. The corresponding Allan deviation was also calculated from this phase noise spectra, shown in the pink trace of Fig. 3 of the main text, and reproduced here in Fig. S4(b). Fractional frequency instabilities of optical-to-microwave conversion and single measurement chain were $5\times10^{-17}$ and $1\times10^{-16}$ at 1 s, respectively. When measuring the absolute stability of the 10 GHz microwaves derived from independent Yb clocks, the noise floor from a single measurement chain was sufficiently low, rendering cross-correlation unnecessary. For example, Allan deviation of 100 s was $3\times10^{-18}$ for our OFD, and $2\times10^{-17}$ for the Yb clocks. In addition, the SDR artifact was shifted to a higher frequency (12 Hz), which is averaged out at time scales exceeding 10 s. We note that in the single chain measurement, amplification of an optically derived 10 GHz tone to +10 dBm is possible while still supporting a fractional frequency instability of $10^{-16}$ at 1 s.

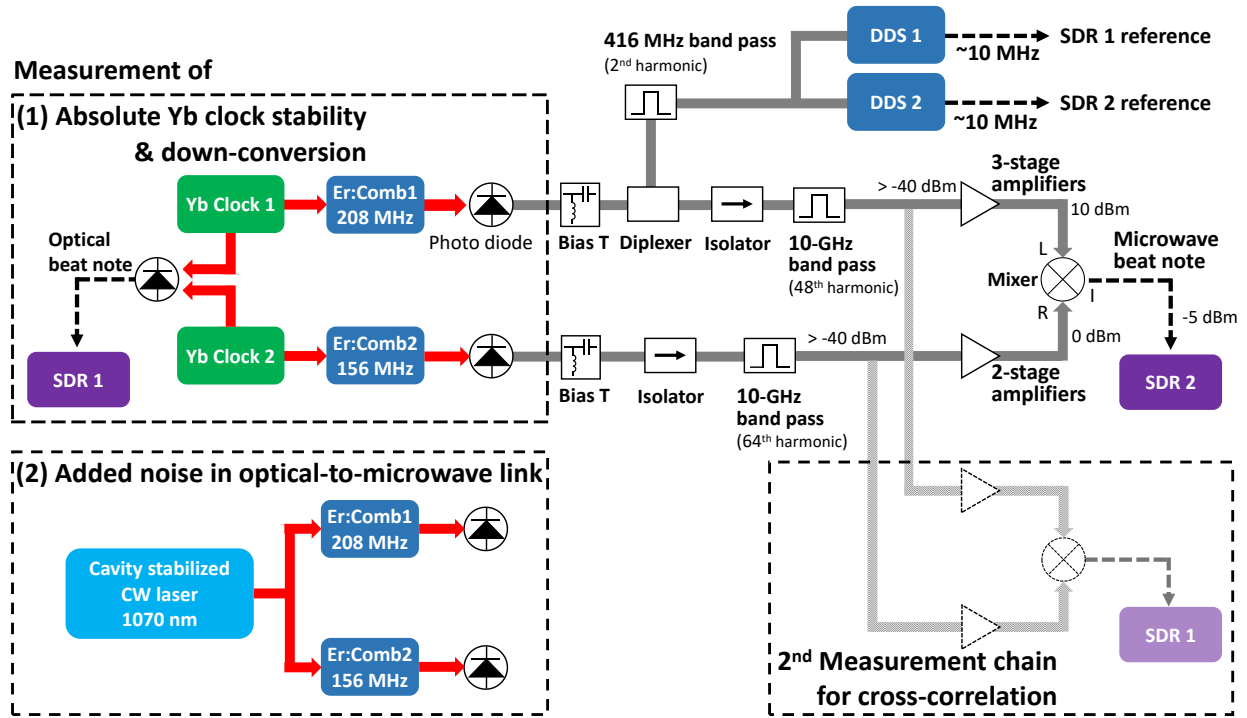

**Fig.S3.** Schematic diagram of the microwave phase and frequency stability measurements. The two microwave signals from the photodiodes are filtered with 10 GHz bandpass filters and amplified to a suitable level for mixer operation. Power levels are as indicated. For measurements when the combs are stabilized to the Yb clocks, we used a single measurement chain (setup(1)). For measurement of the optical-to-microwave conversion noise floor, we used two chains and cross-correlation (setup(2)). The 2nd harmonic of the repetition rate of comb1 was used as a reference signal for the direct digital synthesizers (DDS). The output signals of the DDS's were used as references for software defined radios (SDR).

Phase fluctuation of the 10 GHz signals, Yb clocks and point-by-point difference were measured simultaneously as shown in Fig. 2(a) in the main text, and corresponding fractional frequency instability was also calculated as shown in Fig. S5. The resulting stability of point-by-point difference is consistent with that of single chain measurement. An instability of $1\times10^{-19}$ was obtained at 20,000 s averaging. It was used for the frequency accuracy evaluation (table 1) in the main text.

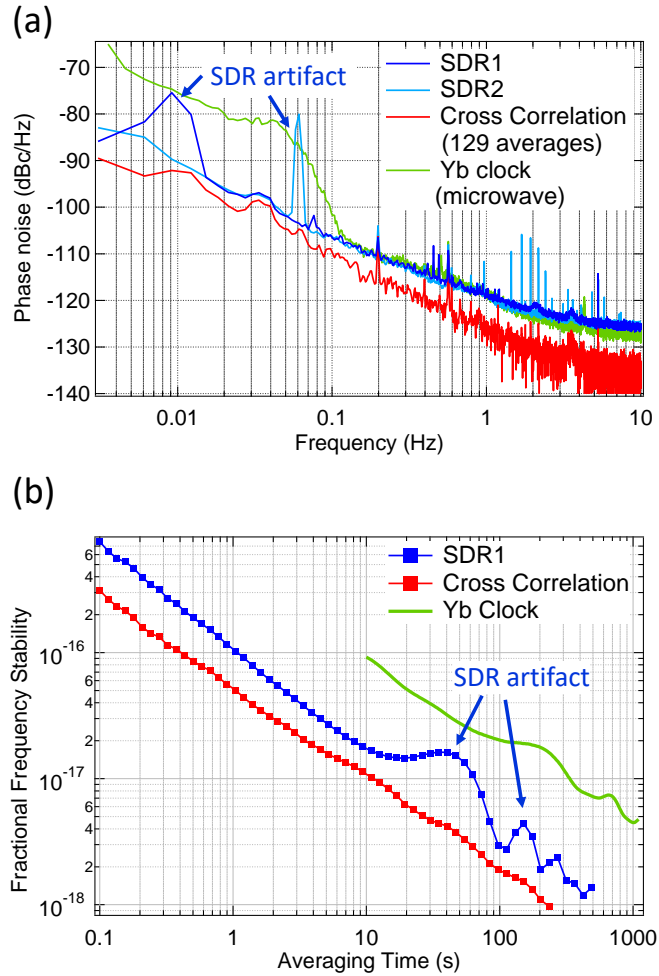

**Fig.S4.** (a) Phase noise of the optical-to-microwave link measured by the individual measurement chains (blue) and from the cross correlation of two chains (red). With 129 averages, the phase noise was suppressed 5-10 dB. Green line shows phase noise of 10 GHz signals when it was stabilized to Yb clocks. (b) Corresponding Allan deviation plots. For comparison, green plot shows the stability of Yb clocks reported in the main text. Cross correlation (red) achieved $5\times10^{-17}$ at 1 s. The result of single measurement chain (blue) is sufficiently lower than Yb clocks.

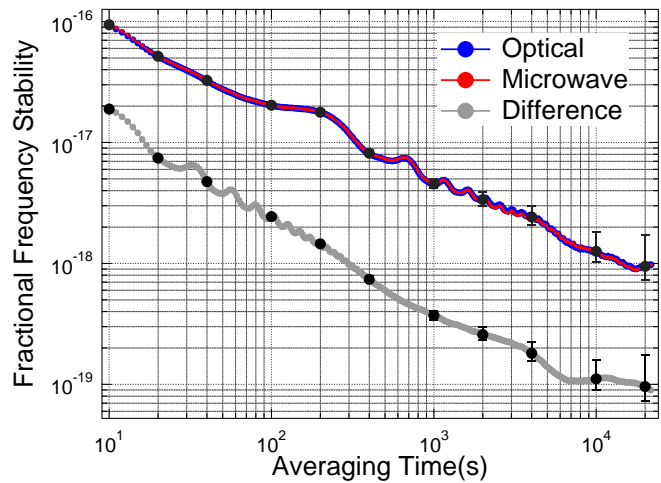

**Fig.S5.** Allan deviation plots of Yb clock (blue), 10 GHz microwave (red), and point-by-point difference (gray). Error bars indicate the 1σ confidence interval calculated by total Allan deviation.

Impact of the optically derived reference on the measured microwave stability

In our microwave measurement, one of the microwave signals and the SDR reference were both derived from the same Yb clock. Therefore, there is a possibility that the noise of our microwaves is underestimated due to common-mode rejection of the Yb clock laser noise. Here we show this common-mode rejection had a negligible impact on our measurements. Figure S6 shows the details of the microwave measurement scheme important for this analysis.

Phase fluctuations of Yb clocks 1 and 2 can be defined as $\varphi_{yb1}(t)$ and $\varphi_{yb2}(t)$, respectively. Since here we are interested in estimating the impact of common-mode rejection, we can assume phase locking is perfect. This leads to phase fluctuations of the repetition rates of the frequency combs $\varphi_{comb1}(t)$ and $\varphi_{comb2}(t)$ of

$$\varphi_{comb1}(t) = \frac{f_{rep1}}{\nu_{yb1}} \times \varphi_{yb1}(t) \tag{1}$$

$$\varphi_{comb2}(t) = \frac{f_{rep2}}{\nu_{yb2}} \times \varphi_{yb2}(t). \tag{2}$$

Here, $f_{rep1}$ and $f_{rep2}$ are the repetition frequencies of each comb and $\nu_{yb1}$ and $\nu_{yb2}$ are optical frequencies of the Yb clocks. The phase fluctuation of the 10 GHz signals, $\varphi_{10GHzcomb1}(t)$ and $\varphi_{10GHzcomb2}(t)$, (including photodiode noise) are given by

$$\varphi_{10GHzcomb1}(t) = 48 \times \frac{f_{rep1}}{\nu_{yb1}} \times \varphi_{yb1}(t) + \varphi_{PD1}(t) \tag{3}$$

$$\varphi_{10GHzcomb2}(t) = 64 \times \frac{f_{rep2}}{\nu_{yb2}} \times \varphi_{yb2}(t) + \varphi_{PD2}(t), \tag{4}$$

where $\varphi_{PD1}(t)$ and $\varphi_{PD2}(t)$ are phase fluctuations originating from the photodiodes. Similarly, phase fluctuation of the 416-MHz signals, $\varphi_{416MHzcomb1}(t)$, used to clock the DDS (including photodiode noise), is given by

$$\varphi_{416MHzcomb1}(t) = 2 \times \frac{f_{rep1}}{\nu_{yb1}} \times \varphi_{yb1}(t) + \varphi_{PD1}(t). \tag{5}$$

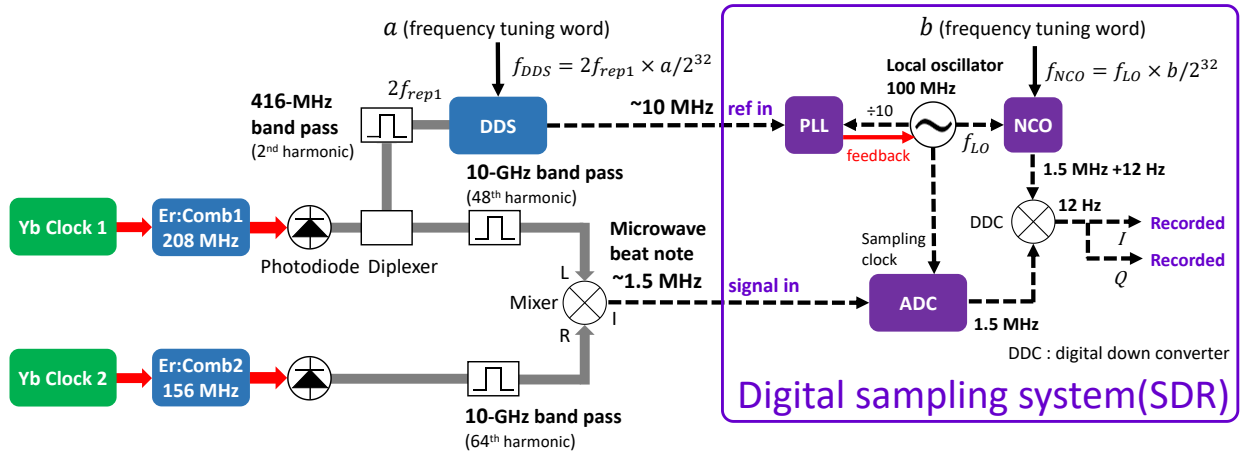

**Fig.S6.** Schematic diagram of microwave measurement with more details in the SDR operation. The reference signal is ultimately down-converted to approximately 1.5 MHz, and is digitally mixed with the signal input. The down-converted 12 Hz signal was recorded through IQ demodulation. The local oscillator in the SDR was also used as the sampling clock for the ADC. ADC, analog-to-digital convertor; SDR, software defined radio, DDC, digital-down converter.

Here, we have assumed the worst case, where that photodiode noise does not scale with carrier frequency, but rather is just as large for the 2$^{nd}$ harmonic as it is for the 48$^{th}$ harmonic.

The resulting phase fluctuations of 10-MHz DDS output due to noise on the input clock, $\varphi_{DDS}(t)$, is

$$\varphi_{DDS}(t) = \frac{10}{416} \times 2 \times \frac{f_{rep1}}{\nu_{yb1}} \times \varphi_{yb1}(t) + \frac{10}{416} \times \varphi_{PD1}(t), \tag{6}$$

the phase fluctuation of 1.5-MHz NCO output, $\varphi_{NCO}(t)$, is

$$\varphi_{NCO}(t) = \frac{1.5}{208} \times \frac{f_{rep1}}{\nu_{yb1}} \times \varphi_{yb1}(t) + \frac{1.5}{416} \times \varphi_{PD1}(t), \tag{7}$$

and the phase fluctuation of microwave beat note (1.5 MHz), $\varphi_{signal}(t)$, is

$$\varphi_{signal}(t) = \varphi_{10GHzcomb1}(t) - \varphi_{10GHzcomb2}(t)$$
$$= 48 \times \frac{f_{rep1}}{\nu_{yb1}} \times \varphi_{yb1}(t) + \varphi_{PD1}(t) - 64 \times \frac{f_{rep2}}{\nu_{yb2}} \times \varphi_{yb2}(t) + \varphi_{PD2}(t). \tag{8}$$

Finally, the relative phase fluctuation of beat note between NCO and microwave, $\varphi_{beat}(t)$, is

$$\varphi_{beat}(t) = \varphi_{signal}(t) - \varphi_{NCO}(t)$$
$$= \left(48 - \frac{1.5}{208}\right) \times \frac{f_{rep1}}{\nu_{yb1}} \times \varphi_{yb1}(t) + \left(1 - \frac{1.5}{416}\right) \varphi_{PD1}(t)$$
$$-64 \times \frac{f_{rep2}}{\nu_{yb2}} \times \varphi_{yb2}(t) + \varphi_{PD2}(t) \tag{9}$$

By comparing this expression to one without any common mode rejection, the Yb clock phase noise reduction is determined to be

$$10\log\left(\frac{48 - \frac{1.5}{208}}{48}\right)^2 = 10\log(1 - 1.5 \times 10^{-4})^2 = -0.0013 \, dB \tag{10}$$

while the photodiode phase noise reduction is given by

$$10\log\left(1 - \frac{1.5}{416}\right)^2 = 10\log(1 - 0.036)^2 = -0.31 \, dB \tag{11}$$

Again, we stress that we have assumed the worst case where the photodiode noise in correlated among all harmonics of the electrical pulse train. In terms of fractional frequency stability, the phase noise reduction would result in a maximum shift in the second significant digit of the Allan deviation plot, virtually indistinguishable from the data presented in Fig. 3 of the main article.

Optical-to-microwave transfer accuracy calculation

The relationship between the frequencies of the Yb clock lasers and frequency combs is given by

$$\nu_{yb1,2} = f_{0_{1,2}} + n \times f_{rep1,2} + f_{beat1,2} \tag{12}$$

where $\nu_{yb1,2}$ are the frequency of the Yb clocks, $f_{0_{1,2}}$ are the OFC offset frequencies, $f_{rep1,2}$ are the OFC repetition rates, and $f_{beat1,2}$ are the frequency difference between the Yb clock and the nearest comb mode of the OFC. Note that the value of $n$ will be different for the two combs. Since both combs use single-actuator feedback by mixing the offset frequency with the optical beat, we may write a single $f_{beat}$ term, set to equal 126.5 MHz for both combs in our experiment, as

$$f_{beat} = f_{0_1} + f_{beat1} = f_{0_2} + f_{beat2} \tag{13}$$

The microwave synthesizer used to generate $f_{beat}$ was referenced to a H-maser. Measurements were performed on the 48[th] and 64[th] harmonics, respectively, of comb 1 and comb 2. These signals, both near 10 GHz, were mixed, producing a difference frequency $f_{sig}$ of

$$f_{sig} = 48 \times f_{rep1} - 64 \times f_{rep2} = 48 \times \left(\frac{\nu_{yb1} - f_{beat}}{n}\right) - 64 \times \left(\frac{\nu_{yb2} - f_{beat}}{m}\right). \tag{14}$$

Here, $n$ and $m$ are the mode number of comb1 and 2, respectively. The nominal value of $f_{sig}$ was 1.5 MHz. This signal was digitized by the SDR, and compared to the output of a digitized and frequency-divided version of the DDS output. The output frequency of a DDS is determined by the input clock frequency, the number of bits in the phase register, and the frequency tuning word (an integer) derived from user input. The DDS's used in our experiment used a 32-bit phase register and was clocked at twice the repetition rate of one of the frequency combs, yielding an output frequency of

$$f_{DDS} = \frac{FTW_{DDS}}{2^{32}} \times 2 \times f_{rep1} = \frac{FTW_{DDS}}{2^{32}} \times 2 \times \left(\frac{\nu_{yb1} - f_{beat}}{n}\right) \tag{15}$$

where $FTW_{DDS}$ is the frequency tuning word of the DDS. The DDS frequency tuning word was set to produce an output near 10 MHz to clock the SDR. The SDR further synthesizes the DDS input to create $f_{NCO}$ near 1.5 MHz. The SDR first multiplies $f_{DDS}$ by 10 to phase-lock its internal 100 MHz quartz oscillator. The numerically controlled oscillator frequency of the SDR was then set close to 1.5 MHz by inputting the correct tuning word, yielding

$$f_{NCO} = \frac{FTW_{NCO}}{2^{32}} \times 10 \times f_{DDS} = \frac{FTW_{NCO}}{2^{32}} \times 10 \times \frac{FTW_{DDS}}{2^{32}} \times 2 \times \left(\frac{\nu_{yb1} - f_{beat}}{n}\right) \tag{16}$$

Comparing $f_{sig}$ to $f_{NCO}$ results in the difference frequency $f_m$, expressed as

$$f_m = \frac{FTW_{NCO}}{2^{32}} \times 10 \times \frac{FTW_{DDS}}{2^{32}} \times 2 \times \left(\frac{v_{yb1} - f_{beat}}{n}\right)$$
$$- \left(48 \times \left(\frac{v_{yb1} - f_{beat}}{n}\right) - 64 \times \left(\frac{v_{yb2} - f_{beat}}{m}\right)\right) \quad (17)$$

The frequency offset between the Yb clocks is defined as $v_{offset} = v_{yb2} - v_{yb1}$. Rewriting $f_m$ in terms of $v_{offset}$ yields

$$f_m = \frac{FTW_{NCO}}{2^{32}} \times 10 \times \frac{FTW_{DDS}}{2^{32}} \times 2 \times \left(\frac{v_{yb1} - f_{beat}}{n}\right)$$
$$- \left(\left(\frac{48}{n} - \frac{64}{m}\right)(v_{yb1} - f_{beat}) - \left(\frac{64}{m}\right)v_{offset}\right) \quad (18)$$

Solving for $v_{offset}$:

$$v_{offset} = \frac{m}{64} \times \left(f_m + \left(\left(\frac{48}{n} - \frac{64}{m}\right) - \frac{FTW_{NCO}}{2^{32}} \times 10 \times \frac{FTW_{DDS}}{2^{32}} \times \frac{2}{n}\right) \times (v_{yb1} - f_{beat})\right) \quad (19)$$

This can be compared to the $v_{offset}$ measured optically after a few small corrections. First, the frequency $f_{beat}$ must be corrected to account for the frequency offset of the H-maser. Second, because the SDR ADC sampling clock is derived from the DDS, $f_m$ must be corrected to account for the fact that $f_{DDS}$ is not at exactly 10 MHz. Finally, in order to measure the optical beat between the Yb clock frequencies, an AOM was used to shift one of the clock outputs by 1.5 MHz. Subtracting this shift and applying the other corrections mentioned above yields

$$v_{offset} = \frac{m}{64} \times f_m \times \frac{FTW_{DDS}}{2^{32}} \times 2 \times \left(\frac{v_{yb1} - f_{beat}}{n}\right) \times \frac{1}{10\ MHz}$$
$$+ \frac{m}{64} \times \left(\left(\frac{48}{n} - \frac{64}{m}\right) - \frac{FTW_{NCO}}{2^{32}} \times 10 \times \frac{FTW_{DDS}}{2^{32}} \times \frac{2}{n}\right)$$
$$\times (v_{yb1} - f_{beat} \times \text{Hmaser shift}) - 1.5\ MHz \times \text{Hmaser shift} \quad (20)$$

Table S1 gives the values of the various parameters used in the calculation. The resulting value of $v_{offset}$ is given in Table 1 of the main text.

| Parameters | Values |
|---|---|
| Absolute frequency of Yb1 $\nu_{yb1}$ (including all AOM shifts) | 259147886795431.25 Hz ($\pm 2.1 \times 10^{-16}$) |
| $f_{beat}$ | 126.5 MHz |
| Hydrogen maser shift | $1 - 6.46 \times 10^{-13}$ (fractional) |
| Mode number of comb1 ($n$) | 1243788 |
| Mode number of comb2 ($m$) | 1658615 |
| Frequency tuning word of DDS ($FTW_{DDS}$) | 103069168 |
| Frequency tuning word of SDR ($FTW_{NCO}$) | 59821114 |
| difference frequency ($f_m$) | -12.694921924988 Hz |
| $\nu_{offset}$ (from Eq. 20) | -0.0152926 Hz |

**Table.S1.** Parameters for optical-to-microwave transfer accuracy calculation. The absolute frequency of the Yb clock was measured against the SI second (*24*). This value includes all frequency shifts between the optical clock and the comb, such as the shift caused by the AOM for fiber noise cancellation. For mode number determination, the repetition rates were measured by a frequency counter and the absolute frequency of the Yb clock was used. Frequency tuning words were determined by software for the direct digital synthesizer (DDS) and the software defined radio (SDR). The difference frequency $f_m$ is the value measured by the SDR after 44,000 seconds of averaging.

Frequency errors in the synthesis chain were examined to ensure no systematic frequency shifts occurred at the $1 \times 10^{-19}$ level. In addition to the H-maser shift mentioned above, effects of phase truncation in the DDS and CORDIC phase estimation in the SDR were evaluated. In order to reduce the required memory, a DDS typically truncates the number of phase bits when transferring to the phase-to-voltage look-up table (*44*). For a DDS with an n-bit phase register that is truncated to m-bits, the maximum possible phase error is $2\pi(1/2^m - 1/2^n)$. This leads to a maximum fractional frequency offset (phase slope difference) given by,

$$\frac{maximum\ phase\ error}{total\ accumulated\ phase} = \frac{\frac{1}{2^m} - \frac{1}{2^n}}{\frac{FTW}{2^n} \times f_{in} \times t} = \frac{\frac{1}{2^m} - \frac{1}{2^n}}{f_{out} \times t} \quad (21)$$

Here, $FTW$ is the frequency tuning word, $f_{in}$ is the input frequency, $f_{out}$ is the output frequency, and $t$ is the time. In our case, the phase was truncated from 32 bits to 16 bits. The maximum fractional frequency offset due to this truncation is shown in Fig. S7(a). As indicated in Eq. 21, the offset decreases as 1/t. The blue line shows the offset contribution to our 10 GHz measurement, where the benefit from translating the noise on the 10 GHz carriers to a 1.5 MHz carrier is apparent. For measurement times longer than $10^4$ s, this effect is much smaller than $10^{-19}$, and can be ignored in our microwave uncertainty calculation. We stress this is the worst-case estimation. Furthermore, we note that a slower sampling rate may average the first and last points of the phase, which leads to greater suppression of the phase slope error.

In the case of the NCO in the SDR, instead of look-up table, COordinate Rotation DIgital Computer (CORDIC) phase estimation is implemented (*45*). CORDIC is a simple algorithm to calculate trigonometric functions. The precision of CORDIC is scaled by the number of the calculation stages. Angle resolution at stage K is given by $\tan^{-1}(1/2^K)$.

In our case, the SDR has a 20-stage CORDIC, leading to an angle resolution estimation of $9.54 \times 10^{-7}$ radians. After the CORDIC calculation, the angle was numerically rounded to 24 bits. However, the 24-bit rounding has higher resolution than 20-stage CORDIC. Therefore, the phase error at the CORDIC was dominant. In a similar way to the DSS phase truncation, the maximum fractional frequency offset was calculated as shown in Fig.S7(b). Here, $f_{out}$ was 1.5 MHz. The effect on our 10 GHz measurement, shown in the blue curve, was also negligible for our uncertainty calculation.

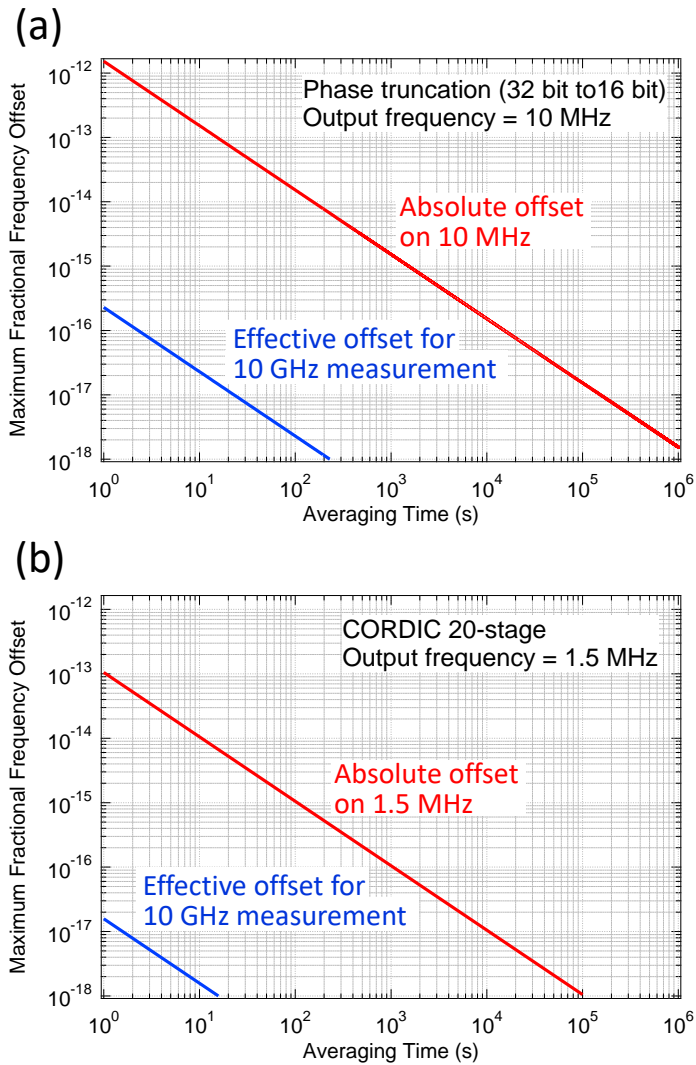

**Fig.S7.** Averaging time dependence of the maximum fractional offset incurred by phase truncation. Red lines show absolute frequency offset on carrier frequencies. Blue lines show the effective offset for 10 GHz measurement. (a) Frequency offset due to phase truncation from 32 bit to 16 bit in the DDS. (b) Frequency offset due to phase truncation from 32 bit to 20-stage CORDIC in the SDR.